\documentclass[12pt]{article}
\usepackage{fancyhdr}
\usepackage{isomath}
\usepackage{amsmath}
\usepackage{amsbsy}
\usepackage{amssymb}
\usepackage{amscd}
\usepackage{amsfonts}
\usepackage{verbatim}
\usepackage{euscript}
\usepackage{alltt}
\usepackage{graphicx,color}
\usepackage[font=small,format=plain,labelfont=bf,up,textfont=it,up]{caption}
\usepackage{subfigure}  
\usepackage[margin=1in]{geometry}

\usepackage{amsmath}
\usepackage{amsbsy}
\usepackage{amssymb}
\usepackage{amscd}
\usepackage{amsfonts}

\newcommand{\bfe}{{\mathbold e}}

\newcommand{\bfC}{{\mathbold C}}

\newcommand{\bfF}{{\mathbold F}}

\newcommand{\bfR}{{\mathbold R}}

\newcommand{\bfU}{{\mathbold U}}

\newcommand{\beq}{\begin{equation}}
\newcommand{\eeq}{\end{equation}}
\newcommand{\beqs}{\begin{eqnarray}}
\newcommand{\eeqs}{\end{eqnarray}}
\newcommand{\beql}{\begin{equation} \label}

\newcommand{\bfalpha}{\mathbold{\alpha}}
\newcommand{\bfomega}{\mathbold{\omega}}

\newcommand{\curl}{\mathop{\rm curl}\nolimits}

\newcommand{\del}{\ensuremath{\partial}}

\begin{document}
\title{An observation on the experimental measurement of dislocation density}
\author{ Amit Acharya\thanks{Corresponding Author: Civil \& Environmental Engineering, Carnegie Mellon University, Pittsburgh, PA 15213, USA; email - acharyaamit@cmu.edu.} \ \ and \ Robin J. Knops\thanks{The Maxwell Institute of Mathematical Sciences and School of Mathematical and Computing Sciences, Heriot-Watt University, Edinburgh, EH14 4AS, Scotland, UK, email - R.J.Knops@hw.ac.uk.} \\ \\
}

\maketitle
\begin{abstract}
\noindent The common practice of ignoring the elastic strain gradient in measurements of geometrically necessary dislocation (GND) density is critically examined. It is concluded that the practice may result in substantial errors. Our analysis points to the importance of spatial variations of the elastic strain field in relation to its magnitude in inferring estimates of dislocation density from measurements.

\end{abstract}

\section{The role of elastic strain gradients in measurement of ``geometrically-necessary'' dislocation (GND) density}

In the standard operating procedure for measuring (and thinking about) the mesoscale dislocation density tensor field, it is generally assumed that the $curl$ of the elastic strain may be ignored when the elastic strain field is of small magnitude in a plastically deformed material. This belief is widely held among researchers in materials science and mechanics working on plasticity and residual stress. For example, a recent published assessment concludes that a small lattice strain and its associated gradient field may be neglected in the evaluation of the  dislocation density, whenever the rotation field is non-uniform. In this short note, we demonstrate circumstances in which this conclusion is erroneous. Our discussion depends upon the fact that a function of small magnitude may not possess a derivative that is of equally small magnitude.

The main argument is the following. Consider a skew-symmetric tensor field $\bfomega(x_1,x_2,x_3)$, a symmetric tensor field $\bfe(x_1,x_2,x_3)$ and a small nondimensional parameter $0 < \nu \ll 1$. We assume that the ratio of the magnitude of $curl\,\bfe$ to that of $curl\,\bfomega$ is approximately unity or less. We define the elastic distortion $\bfU^e$ as
\[
\bfU^e := \bfomega + \nu \bfe.
\]
The \emph{Kr\"{o}ner-Nye} dislocation density field, defined as
 \[
 \bfalpha = curl\, \bfU^e 
 \]
may be approximated by $curl\,\bfomega$ provided that $\nu curl\, \bfe$ is not of a higher order of magnitude than $curl\,\bfomega$. For example, let the only non-vanishing components of $\bfomega$ be $\omega_{12}= -\omega_{21} = k x_1$. Also, let the only non-vanishing component of $\bfe$ be $e_{11} = k x_2$. Then, the magnitudes $|curl\,\bfomega| = k$, $|curl\,\bfe| = k$, and the only non-vanishing component of $\bfalpha = curl\,\bfomega + curl\,(\nu \bfe)$ is $\alpha_{13} = k - \nu k \approxeq k$. However, it is equally clear that if the ratio $|\curl\,\bfe| / |curl\,\bfomega|$ were to be large (e.g. $\approxeq 1/\nu$), then such an approximation does not hold. In what follows, we illustrate these remarks with explicit counterexamples.

\section{Micro, and meso-macro scalings}

For simplicity, we consider simply-connected domains for all arguments that follow. Consider a symmetric tensor field $\epsilon_{ij}(y_1,y_2,y_3)$ as a function of the rectangular Cartesian coordinates $(y_i, i = 1 \,\text{to} \, 3)$, where the components of $\epsilon_{ij}$ are also expressed with respect to the basis of the same coordinate system (as well as all tensor components below). Let us \emph{assume} that the tensor field is twice-differentiable and bounded, and, without loss of generality, that the maximum magnitude of any of the components of the $\epsilon_{ij}$ field does not exceed unity for all values of $(y_1,y_2,y_3)$. The field is also assumed to be compatible in the sense of satisfying the Saint-Venant compatibility conditions, i.e. it can be expressed as the symmetrized gradient (or the strain tensor) of some vector field (the corresponding displacement field):
\begin{equation}\label{y-compatibility}
\frac{\del^2 \epsilon_{il}}{\del y_k \del y_m} - \frac{\del^2 \epsilon_{kl}}{\del y_i \del y_m} - \frac{\del^2 \epsilon_{im}}{\del y_k \del y_l} + \frac{\del^2 \epsilon_{km}}{\del y_i \del y_l} = 0.
\end{equation}
We regard this field as an elastic strain field viewed at the microscopic scale. The terms micro, meso, macro here refer to the length-scale of variation, and therefore of measurements, of various fields. Let us define a \emph{small-magnitude}, compatible elastic strain field $e_{ij}$ at the meso/macroscopic scale according the following definition:
\begin{equation}\label{e-field}
e_{ij}(x_1,x_2,x_3) := \nu \epsilon_{ij}\left(\frac{x_1}{\nu^m},\frac{x_2}{\nu^m},\frac{x_3}{\nu^m}\right),
\end{equation}
where $0 < \nu \ll 1$ is the same non-dimensional scale factor introduced previously and $ m \geq 2$ is, say, an integer. We use the scaling
\[
y_i = \frac{x_i}{\nu^m};
\]
consequently, $y_i$ is a fine (magnified) length scale of observation that resolves the microscopic fields of, say, wave-length $l$, and $x_i$ is a coarse length scale of observation resolving meso-macro fields of wave-length $L \gg l$. We think of $\nu = l/L$.

It is now easy to see that
\[
\frac{\del^2 e_{il}}{\del x_k \del x_m} - \frac{\del^2 e_{kl}}{\del x_i \del x_m} - \frac{\del^2 e_{im}}{\del x_k \del x_l} + \frac{\del^2 e_{km}}{\del x_i \del x_l} = \frac{1}{\nu^{2m-1}} \left( \frac{\del^2 \epsilon_{il}}{\del y_k \del y_m} - \frac{\del^2 \epsilon_{kl}}{\del y_i \del y_m} - \frac{\del^2 \epsilon_{im}}{\del y_k \del y_l} + \frac{\del^2 \epsilon_{km}}{\del y_i \del y_l} \right) = 0.
\]
Thus, the elastic strain field $e_{ij}$ is compatible, meaning that there exists a displacement field $u_i(x_1,x_2,x_3)$ such that
\[
\frac{1}{2} \left( \frac{\del u_i}{\del x_j} + \frac{\del u_j}{\del x_i} \right) = e_{ij}.
\]
But then, defining the elastic rotation field as
\[
\frac{1}{2} \left( \frac{\del u_i}{\del x_j} - \frac{\del u_j}{\del x_i} \right) = \omega_{ij},
\]
we note that the $\alpha_{ij}$ field, corresponding to the elastic distortion field $u_{i,j}$, vanishes but the $curl$ of the elastic rotation field is in general large:
\[
0 = \alpha_{il} = \varepsilon_{lkj} \frac{\del^2 u_i}{\del x_j \del x_k} = \varepsilon_{lkj} \frac{\del \omega_{ij}}{\del x_k} + \varepsilon_{lkj}\frac{\del e_{ij}}{\del x_k}
\]
\[
\Rightarrow \varepsilon_{lkj} \frac{\del \omega_{ij}}{\del x_k}(x_1,x_2,x_3) = - \frac{1}{\nu^{m-1}} \varepsilon_{lkj} \frac{\del \epsilon_{ij}}{\del y_k}\left(\frac{x_1}{\nu^m},\frac{x_2}{\nu^m},\frac{x_3}{\nu^m}\right),
\]
where $\varepsilon_{ijk}$ are components of the alternating tensor. An example of a displacement field $\tilde u_i$ whose displacement gradient obviously has vanishing $curl$ but whose strain and rotation fields do not is given by $\tilde{u}_1 = k x_2 x_1$, $\tilde{u}_2 = -k x_1^2/2$, $\tilde{u}_3 = 0$, $k$ a constant. Then, in an obvious notation, $\tilde{e}_{11} = kx_2$ and all other strain components are zero, while $\tilde{\omega}_{12} = kx_1$ and $\tilde{\omega}_{13} = \tilde{\omega}_{23} = 0$.

In the next Section, we explicitly construct strain ($\epsilon_{ij}$) and rotation ($\omega_{ij}$) fields that provide counterexamples to the claim that small strains necessarily mean that the strain field can be neglected in the determination of the dislocation density field.

\section{Counterexamples}

For the considerations of the previous section to be relevant, it remains to be shown that strain fields $\epsilon_{ij}$ satisfying (\ref{y-compatibility}) in fact exist. But this, of course, is a simple matter. Consider \emph{any} smooth displacement field $v_i(y_1,y_2,y_3)$ on a domain, viewed at a resolution of, say, a tenth of a micron. By this we mean that when $y_i$ are plotted on a scale of a tenth of a micron, an unambiguous specification of $v_i$ as a function of these coordinates can be made. We require at least that some of the components of the $v_i$'s third-order spatial derivatives do not vanish. For example, we could take each $v_i, i = 1 \, \text{to} \, 3$, to be a fourth degree polynomial in $(y_1,y_2,y_3)$ or trigonometric functions. We then define
\[
\epsilon_{ij}(y_1,y_2,y_3) := \frac{1}{2} \left( \frac{\del v_i}{\del y_j} + \frac{\del v_j}{\del y_i} \right) (y_1,y_2,y_3).
\]
By construction, any such field $\epsilon_{ij}$ is guaranteed to satisfy (\ref{y-compatibility}). In terms of this $\epsilon_{ij}(y_1,y_2,y_3)$ field, define the $e_{ij}(x_1,x_2,x_3)$ field from (\ref{e-field}) for $\nu = 0.01$ and $m = 2$. Roughly, this amounts to viewing the same strain field at a scale of a millimeter. Thus, when for example the field $\epsilon_{ij}(y_1,y_2,y_3)$ contains well-resolved sinusoidal oscillations, the field $e_{ij}(x_1,x_2,x_3)$ would appear highly oscillatory. The field $u_i$ of the construction above corresponding to this choice of $e_{ij}$ field would be
\[
u_i(x_1,x_2,x_3) = \nu^3 \,v_i \left( \frac{x_1}{\nu^2},\frac{x_2}{\nu^2},\frac{x_3}{\nu^2} \right).
\]
This construction is an explicit example for which \emph{there is no dislocation density in the body but if approximated by the elastic rotation gradient, can be erroneously calculated as being (arbitrarily) large, depending on the rapidity of the spatial fluctuation of the small-magnitude elastic strain field}. 

In this counterexample, the dislocation density is exactly zero. However, a small magnitude, but rapidly oscillating, elastic strain field at the mesoscale would seem to be a common occurrence in the presence of a large number of dislocations in the body. Elements of the above example show that the $curl$ of the elastic strain in such cases can be large in magnitude so that its neglect can lead to substantial underestimation of the actual (excess/polar/geometrically-necessary) dislocation density. To illustrate this remark,  let the elastic distortion be composed of a rotation tensor  $\bfomega$ with only non-zero components $\omega_{12}= -\omega_{21} = k x_1$ as before, and let the only non-vanishing component of $\bfe$ be $e_{11} = \nu k \sin(x_2/\nu^2)$, which is a small-magnitude, very high frequency oscillation. Then, the only non-vanishing component of $\bfalpha = curl\,\bfomega + curl\,\bfe$ is $\alpha_{13} = k - (k/\nu) \cos(x_2/\nu^2)$, while $|curl\,\bfomega| = k$.

\section{Concluding remarks}

The discussion so far has been confined to linear kinematics, but even in the geometrically nonlinear case we may treat a Right Cauchy Green tensor (RCG) field $\bfC = {\bfF}^T \bfF$ of a smooth deformation field with deformation gradient $\bfF$. Consequently, its Riemann-Christoffel tensor is zero, and its essentially unique rotation field $\bfR$  that also satisfies $\bfF = \bfR \sqrt{\bfC}$ (up to a rigid rotation) in
general is not $curl$-free. Thus, in general, a compatible, i.e. $curl$-free, $\bfF$ field has a non-uniform rotation field with non-vanishing $curl$. In particular, Shield \cite{shield1973rotation} shows that this result holds for compatible finite strain fields of small magnitude with large gradients:
\[
C_{ik} - \delta_{ik} = O(\nu), \ \ \ \ C_{ik,l} = O(1)
\]
and
\[
R_{mn,k} = \frac{1}{2} R_{mp} (C_{pk,n} - C_{nk,p}) + O(\nu).
\]
Of course, the $curl$ of an RCG field arising from a smooth deformation does not necessarily vanish.

A second remark pertains to the fact that experimental estimates of GND density are based on a difference approximation of the derivative where the spatial step-length on the mesoscale does not really tend to zero. One may then ask whether the `discrete' derivative, based on a small, but finite, mesoscale step-length $\Delta x$, of a continuously differentiable function with small magnitude, say,
\[
g(x) = \nu f\left(\frac{x}{\nu^m}\right), 0< \nu \ll 1, m\geq 2,
\]
is small (we assume $f$ to be a bounded by unity, nondimensional, continuously differentiable function). If so, then such a definition would allow the neglect of the small magnitude strain field. But even this proposition appears to be untenable. Defining a discrete derivative $g'(x,\Delta x)$ as
\[
g'(x,\Delta x) = \frac{\nu \left[ f\left(\frac{x}{\nu^m}+\frac{\Delta x}{\nu^m}\right) - f\left(\frac{x}{\nu^m}\right) \right]}{\Delta x} = \frac{\nu \left[ f\left(\frac{x}{\nu^m}+\frac{\Delta x}{\nu^m}\right) - f\left(\frac{x}{\nu^m}\right) \right]}{\frac{\Delta x}{\nu^m}} \frac{1}{\nu^m},
\]
it can be observed from the middle expression that the term multiplying $\nu$ is in general not going to be small in magnitude since, even for $\Delta x$ small, the arguments of $f$ are not close. Hence, $g'(x,\Delta x)$ is not going to be small and cannot be ignored. The extreme right-hand-side expression shows that for $\Delta x \rightarrow 0$, $0 < \nu \ll 1$ fixed, $\lim_{\Delta x \rightarrow 0} g'(x,\Delta x)$ is actually large.

\bibliographystyle{alpha} \bibliography{measurement_disloc_density_bib}

\begin{thebibliography}{Shi73}

\bibitem[Shi73]{shield1973rotation}
R.~T. Shield.
\newblock The rotation associated with large strains.
\newblock {\em SIAM Journal on Applied Mathematics}, 25(3):483--491, 1973.

\end{thebibliography}

\end{document}